\documentclass[10pt]{article}

\usepackage{amsmath}
\usepackage{amssymb}

\usepackage{graphicx}

\usepackage{cite}

\usepackage{color} 

\usepackage[labelfont=bf,labelsep=period,justification=raggedright]{caption}

\makeatletter
\renewcommand{\@biblabel}[1]{\quad#1.}
\makeatother

\date{}

\begin{document}

\begin{flushleft}
{\Large
\textbf{Are random trading strategies more successful than technical ones?}
}
\\
A.E.Biondo$^{1,\ast}$, 
A.Pluchino$^{2}$, 
A.Rapisarda$^{2}$,
D.Helbing$^{3}$
\\
\bf{1} Dipartimento di Economia e Impresa - Universit\'a di Catania, Corso Italia 55, 95129 Catania, Italy
\\
\bf{2} Dipartimento di Fisica e Astronomia, Universit\`a di Catania  and INFN sezione di Catania, 
Via S. Sofia 64, 95123 Catania, Italy
\\
\bf{3} ETH Zurich, Clausiustrasse 50, 8092 Zurich, Switzerland
\\
$\ast$ E-mail: ae.biondo@unict.it
\end{flushleft}

\section*{Abstract}

In this paper we explore  the specific role of randomness in financial markets, inspired by the beneficial role of noise in many physical systems and in previous applications to complex socio-economic systems. After a short  introduction, we study the performance of some of the most used trading strategies in predicting the dynamics 
of financial markets for different international stock exchange indexes, with the goal of comparing them to the performance of a completely random strategy. 
In this respect, historical data for FTSE-UK, FTSE-MIB, DAX, and S$\&$P500  indexes are taken into account for a period of about $15$-$20$ years (since their creation until today).


\section*{Introduction}

In physics, both at the classical and quantum level, many real systems work fine and more efficiently due to the useful role of a random weak noise \cite{Kirkpatrick,Parisi,Gammaitoni,Mantegna,Caruso,Toral}. 
But not only physical systems benefits from disorder. In fact, noise has a great influences on the dynamics of cells, neurons and other biological entities, but also on ecological, geophysical and socio-economic systems. Following this line of research, we have recently investigated how random strategies can help to improve the efficiency of a hierarchical group in order to face the Peter principle\cite{Peter,Pluchino1,Pluchino2} or a public institution such as a  Parliament  \cite{Pluchino3}. Other groups have successfully explored similar strategies  in minority and Parrondo games \cite{Satinover1,Satinover2}, in portfolio performance evaluation \cite{Gilles} and in the context of the continuous double auction \cite{Farmer}. 
\\
Recently Taleb has brilliantly discussed in his successful books  \cite{Taleb, Taleb2} how chance and black swans rule our life, but also economy and financial market behavior beyond our personal and rational expectations or control.  Actually, randomness enters in our everyday life although we hardly recognize it. Therefore, even without being skeptic as much as Taleb, one could easily claim that we often misunderstand phenomena around us and are fooled by apparent connections which are only due to fortuity. Economic systems are unavoidably affected by expectations, both present and past, since agents' beliefs strongly influence their future dynamics. If today a very good expectation emerged about the performance of any security, everyone would try to buy it and this occurrence would imply an increase in its price. Then, tomorrow, this security would be priced higher than today, and this fact would just be the consequence of the market expectation itself. 
This deep dependence on expectations made financial economists try to build mechanisms to predict future assets prices. The aim of this study is precisely to check whether these mechanisms, which will be described in detail in the next sections, are more effective in predicting the market dynamics compared to a completely random strategy. 
\\
In a previous article \cite{biondo-jsp}, motivated also by some intriguing experiments where a child, a chimpanzee and darts were successfully used for remunerative investments  \cite{wiseman,porter}, we already found some evidence in favor of random strategies for the FTSE-UK stock market. Here we will extend this investigation to other financial markets and for new trading strategies. The paper is organized as follows. Section $2$ presents a brief introduction to the debate about predictability in financial markets. In Section $3$ we introduce the financial time series considered in our study and perform a detrended analysis in search for possible correlations of some kind. In Section $4$ we define  the trading strategies used in our simulations while, in Section $5$, we discuss the main results obtained. Finally, in Section $6$, we  draw our conclusions, suggesting also some counterintuitive policy implications.

\section{Expectations and Predictability in Financial Markets}

As Simon \cite{simon} pointed out, individuals assume their decision on the basis of a limited knowledge about their environment and thus face high search costs to obtain needed information. However, normally, they cannot gather all information they should. Therefore, agents act on the basis of \textit{bounded rationality}, which leads to significant biases in the expected utility maximization that they pursue. In contrast, Friedman \cite{friedman} defended the rational agent approach, which considers that the behavior of agents can be best described assuming their rationality, since non-rational agents do not survive competition on the market and are driven out of it. Therefore, neither systematic biases in expected utility, nor bounded rationality can be used to describe agents' behaviors and their expectations. \\
Without any fear of contradiction, one could say that nowadays two main reference models of expectations have been widely established within the economics literature: the adaptive expectations model and the rational expectation model. Here we will not give any formal definition of these paradigms. For our purposes, it is sufficient to recall their rationale. The adaptive expectations model is founded on a somehow weighted series of backward-looking values (so that the expected value of a variable is the result of the combination of its past values). In contrast, the rational expectations model hypothesizes that all agents have access to all the available information and, therefore, know exactly the model that describes the economic system (the expected value of a variable is then the objective prediction provided by theory). These two theories dates back to very relevant contributions, among which we just refer to Friedman \cite{friedman,friedman2}, Phelps \cite{phelps}, and Cagan \cite{cagan} for adaptive expectations (it is however worth to notice that the notion of ``adaptive expectations'' has been first introduced by Arrow and Nerlove \cite{arrow-nerlove}). For rational expectations we refer to Muth \cite{muth}, Lucas \cite{lucas}, and Sargent-Wallace \cite{sargent-wallace}.  
\\
Financial markets are often taken as example for complex dynamics and dangerous volatility. This somehow suggests the idea of unpredictability. Nonetheless, due to the relevant role of those markets in the economic system, a wide body of literature has been developed to obtain some reliable predictions. As a matter of fact, forecasting is the key point of financial markets. Since Fama \cite{fama}, we say a market is efficient if \emph{perfect arbitrage} occurs. This means that the case of inefficiency implies the existence of opportunities for unexploited profits and, of course, traders would immediately operate long or short positions until any further possibility of profit disappears. Jensen \cite{jensen} states precisely that a market is to be considered efficient with respect to an information set if it is impossible to make profits by trading on the basis of that given information set. This is consistent with Malkiel \cite{malkiel}, who argues that an efficient market perfectly reflects all information in determining assets' prices. As the reader can easily understand, the more important part of this definition of efficiency relies on the completeness of the information set. In fact, Fama \cite{fama} distinguishes three forms of market efficiency, according to the degree of completeness of the informative set (namely ``weak'', ``semi-strong'', and ``strong'').  Thus, traders and financial analysts continuously seek to expand their information set to gain the opportunity to choose the best strategy: this process involves agents so much in price fluctuations that, at the end of the day, one could say that their activity is reduced to a systematic guess. The complete globalization of financial markets amplified this process and, eventually, we are experiencing decades of extreme variability and high volatility.
\\ 
Keynes argued, many years ago, that rationality of agents and mass psychology (so-called ``animal spirits'') should not be interpreted as if they were the same thing. The Author introduced the very famous \emph{beauty contest} example to explain the logic underneath financial markets. In his General Theory \cite{keynes} he wrote that \emph{``investment based on genuine long-term expectations is so difficult as to be scarcely practicable. He who attempts it must surely lead much more laborious days and run greater risks than he who tries to guess better than the crowd how the crowd will behave; and, given equal intelligence, he may make more disastrous mistakes."}
In other words, in order to predict the winner of the beauty contest, one should try to interpret the jury's preferred beauty, rather than pay attention on the ideal of objective beauty. In financial markets it is exactly the same thing. It seems impossible to forecast prices of shares without mistakes. The reason is that no investor can know in advance the opinion ``of the jury'', i.e. of a widespread, heterogeneous and very substantial mass of investors that reduces any possible prediction to just a guess.
\\
Despite considerations like these, the so-called Efficient Market Hypothesis (whose main theoretical background is the theory of rational expectations), describes the case of perfectly competitive markets and perfectly rational agents, endowed with all available information, who choose for the best strategies (since otherwise the competitive clearing mechanism would put them out of the market). There is evidence that this interpretation of a fully working \textit{perfect arbitrage} mechanism is not adequate to analyze financial markets as, for example: Cutler \emph{et al.} \cite{cutler}, who shows that large price movements occur even when little or no new information is available; Engle \cite{engle} who reported that price volatility is strongly temporally correlated; Mandelbrot \cite{mandelbrot,mandelbrot2}, Lux \cite{lux2},  Mantegna and Stanley \cite{mantegna-stanley} who argue that short-time fluctuations of prices are non-normal; or \emph{last but not least}, Campbell and Shiller \cite{campbell-shiller} who explain that prices may not accurately reflect rational valuations.
\\
Very interestingly, a plethora of heterogeneous agents models have been introduced in the field of financial literature. In these models, different groups of traders co-exist, with different expectations, influencing each other by means of the consequences of their behaviors. Once again, our discussion cannot be exhaustive  here, but we can fruitfully mention at least contributions by Brock \cite{brock,brock2}, Brock and Hommes \cite{brock-hommes}, Chiarella \cite{chiarella}, Chiarella and He \cite{chiarella-he}, DeGrauwe \emph{et al.} \cite{degrauwe}, Frankel and Froot\cite{frankel-froot}, Lux \cite{lux}, Wang \cite{wang}, and Zeeman \cite{zeeman}. 
\\ 
Part of this literature refers to the approach, called ``adaptive belief systems'', that tries to apply non-linearity and noise to financial market models.  Intrinsic uncertainty about economic fundamentals, along with errors and heterogeneity, leads to the idea that, apart from the fundamental value (i.e. the present discounted value of the expected flows of dividends), share prices fluctuate unpredictably because of phases of either optimism or pessimism according to corresponding phases of uptrend and downtrend that cause market crises. How could this sort of erratic behavior be managed in order to optimize an investment strategy?  In order to explain the very different attitude adopted by agents to choose strategies when trading on financial markets, a distinction is done between \textit{fundamentalists} and \textit{chartists}. The former ones base their expectations about future assets' prices upon market fundamentals and economic factors (i.e. both micro- and macroeconomic variables, such as dividends, earnings, economic growth, unemployment rates, etc). Conversely, the latter ones try to extrapolate trends or statistically relevant characteristics from past series of data, in order to predict future paths of assets prices (also known as technical analysis). 
\\
Given that the interaction of these two groups of agents determines the evolution of the market, we choose here to focus on chartists' behavior (since a qualitative analysis on macroeconomic fundamentals is absolutely subjective and difficult to asses), trying to evaluate the individual investor's \textit{ex-ante} predictive capacity. Assuming the lack of complete information, randomness plays a key role, since efficiency is impossible to be reached. This is particularly important in order to underline that our approach does not rely on any form of the above mentioned Efficient Markets Hypothesis paradigm. More precisely, we are seeking for the answer to the following question: if a trader assumes the lack of complete information through all the market (i.e. the unpredictability of stock prices dynamics \cite{blackscholes,merton,coxingersollross,hullwhite}), would an \textit{ex-ante} random trading strategy perform, on average, as good as well-known trading strategies?  We move from the evidence that, since each agent relies on a different information set in order to build his/her trading strategies, no efficient mechanism can be invoked. Instead, a complex network of self-influencing behavior, due to asymmetric circulation of information, develops its links and generates herd behaviors to follow some signals whose credibility is accepted. \\
Financial crises show that financial markets are not immune to failures. Their periodic success is not \emph{free of charge}: catastrophic events burn enormous values in dollars and the economic systems in severe danger. Are traders so sure that elaborated strategies fit the dynamics of the markets? 
Our simple simulation will perform a comparative analysis of the performance of different trading strategies: our traders will have to predict, day by day, if the market will go up ('bullish' trend) or down ('bearish' trend). Tested strategies are: the Momentum, the RSI, the UPD, the MACD, and a completely Random one. 
\\
Rational expectations theorists would immediately bet that the random strategy would loose the competition as it is not making use of any information but, as we will show, our results are quite surprising.  

\begin{figure}
\begin{center}
\includegraphics[width=3.5in,angle=0]{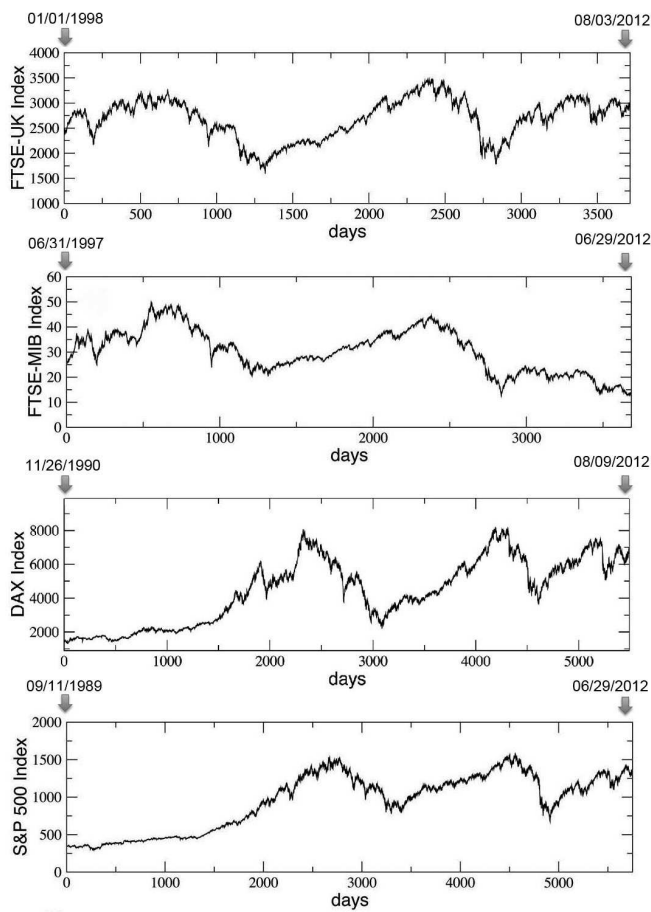}
\end{center}
\caption{{\bf Temporal evolution of four important financial market indexes (over time intervals going from $3714$ to $5750$ days).} From the top to the bottom, we show the FTSE UK All-Share index, the FTSE MIB All-Share index, the DAX All-Share index and the S$\&$P 500 index. See text for further details.}
\end{figure}

\section{Detrended Analysis of the Index Time Series}

We consider four very popular indexes of financial markets 
and in particular, we analyze the following corresponding time series, shown in Fig. 1:   
\\
- {\bf FTSE UK All-Share index}, from January, 1st 1998 to August, 3rd 2012, for a total of $T=3714$ days;
\\
- {\bf FTSE MIB All-Share index}, from December, 31th 1997 to June, 29th 2012, for a total of $T=3684$ days;
\\
- {\bf DAX All-Share index}, from November, 26th 1990 to August, 09th 2012, for a total of $T=5493$ days;
\\
- {\bf S$\&$P 500 index}, from September, 11th 1989 to June, 29th 2012, for a total of $T=5750$ days;
\\
\begin{figure}  
\begin{center}
\includegraphics[width=3.5in,angle=0]{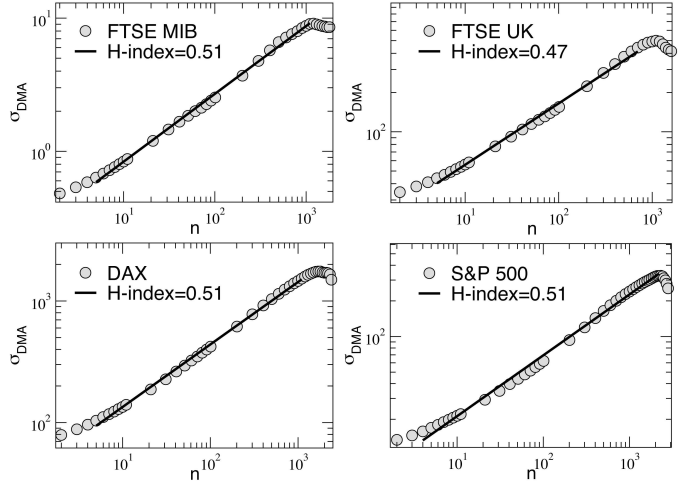}
\end{center}
\caption{{\bf Detrended analysis for the four financial market series shown in Fig.1.} The power law behavior of the DMA standard deviation allows to derive an Hurst index that, in all the four cases, oscillates around $0.5$, thus indicating an absence of correlations, on average, over large time periods. See text.}
\end{figure}
   
In general, the  possibility to predict financial time series has been stimulated by the finding of some kind of persistent behavior in some of them \cite{mantegna-stanley,gabaix,livan}.
The main purpose of the present section is to investigate the possible presence of correlations in the previous four financial series of European and US stock market all share indexes. In this connection, we will calculate the time-dependent Hurst exponent by using the detrended moving average (DMA) technique \cite{Carbone}.  Let us begin with a summary of the DMA algorithm. 
The computational procedure is based on the calculation of the standard deviation $\sigma_{DMA}(n)$ along a given time series defined as  
\begin{equation}
\sigma_{DMA}(n)=\sqrt{\frac{1}{N_{max}-n} \sum_{t=n}^{N_{max}} [y(t)-\tilde{y}_n(t)]^2},
\end{equation}
where  $ \tilde{y}_n(t)=\frac{1}{n} \sum_{k=0}^{n-1} y (t-k)$ is the average calculated in each time window of size $n$.
In order to determine the Hurst exponent $H$, the function $\sigma_{DMA}(n)$ is calculated for increasing values of $n$ inside the interval $[2,N_{max}/2]$, $N_{max}$ being the length of the time series, and the
obtained values are reported as a function of $n$ on a log-log plot. In general, $\sigma_{DMA}(n)$ exhibits a power-law dependence with exponent $H$, i.e. 
\begin{equation}
\sigma_{DMA} \propto  n^H .
\end{equation}
In particular, if $0\le H \le 0.5$, one has a negative correlation or anti-persistent behavior, while if $0.5\le H \le 1$  one has a positive correlation or persistent behavior. The case of $H=0.5$ corresponds to an uncorrelated Brownian process.
In our case, as a first step, we  calculated the Hurst exponent considering the complete series. 
This analysis is illustrated in the four plots of Fig. 2. Here, a linear fit to the log-log plots reveals that all the values of the Hurst index H obtained in this way for the time series studied are, on average, very close to 0.5. This result seems to indicate an absence of correlations on large time scales and a consistence with a random process. 
\begin{figure}  
\begin{center}
\includegraphics[width=3.5in,angle=0]{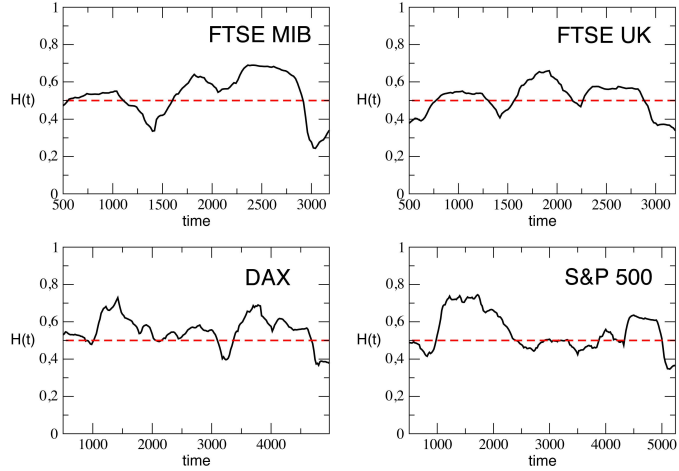}
\end{center}
\caption{{\bf Time dependence of the Hurst index for the four series: on smaller time scales, significant correlations are present. See text.}}
\end{figure}
\\
On the other hand, it is interesting to calculate the Hurst exponent locally in time. In order to perform this analysis, we consider subsets of the complete series by means of sliding windows $W_s$ of size $N_s$, which move along the series with time step $s$. This means that, at each time $t\in[0,N_{max}-s]$, we calculate the $\sigma_{DMA}(n)$ inside the sliding window $W_s$ by changing $N_{max}$ with $N_s$ in Eq.(1).  Hence, following the same procedure described above, a sequence of Hurst exponent values $H(t)$ is obtained as function of time. 
In Fig. 3 we show the results obtained for the parameters  $N_s=1000$, $s=20$.
In this case, the values obtained for the Hurst exponent $H(t)$ differ very much locally from 0.5, thus indicating the presence of 
significant local correlations. 
\\
This investigation, which is in line with what was found previously in Ref. \cite{Carbone} for the Dax index, seems to suggest that correlations are important only on a local temporal scale, while they cancel out averaging over long-term periods. As we will see in the next sections, this feature will affect the performances of the trading strategies considered.

\section{Trading strategies description}

In the present study we consider five trading strategies defined  as  follows:

{\it 1) Random (RND) Strategy}
\\
This  strategy is the simplest one, since the correspondent trader makes his/her  prediction at  time $t$ completely at random (with uniform distribution). 

{\it 2) Momentum (MOM) Strategy}
\\
This strategy is based on  the so called 'momentum' $M(t)$ indicator, i.e. the difference between the value $I(t)$ and the value $I(t-\tau_M)$, where $\tau_M$ is a given trading interval (in days). Then, if $M(t)=I(t)-I(t-\tau_M)>0$, the trader predicts an increment of the closing index for the next day (i.e. it predicts that $I(t+1)-I(t)>0$) and vice-versa. In the following simulations we will consider $\tau_M=7$ days, since this is one of the most used time lag for the momentum indicator. See Ref. \cite{Murphy}.

\begin{figure}  
\begin{center}
\includegraphics[width=2in,angle=0]{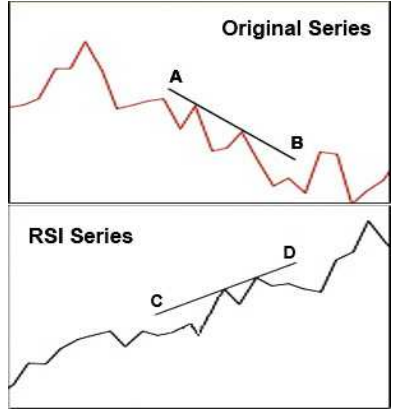}
\end{center}
\caption{{\bf RSI divergence example}. A divergence is a disagreement between the indicator (RSI) and the underlying price. By means of trend-lines, the analyst check that slopes of both series agree. When the divergence occurs, an inversion of the price dynamic is expected. In the example a bullish period is expected.}
\end{figure}

{\it 3) Relative Strength Index (RSI) Strategy}
\\
This strategy is based on a more complex indicator  called 'RSI' . It is considered a measure of the stock's recent trading strength and its definition is: $RSI(t)=100-100/[1+RS(t)]$, where $RS(t,\tau_{RSI})$ is the ratio between the sum of the positive returns and the sum of the negative returns occurred during the last $\tau_{RSI}$ days before $t$. Once calculated the RSI index for all the days included in a given time-window of length $T_{RSI}$ immediately preceding the time $t$, the trader which follows the RSI strategy makes his/her prediction on the basis of a possible reversal of the market trend, revealed by the so called 'divergence' between the original time series and the new RSI one. A divergence can be defined referring to a comparison between the original data series and the generated RSI-series, and it is the most significant trading signal delivered by any oscillator-style indicator. It is the case when the significant trend between two local extrema shown by the RSI trend is oriented in the opposite direction to the significant trend between two extrema (in the same time lag) shown by the original series. When the RSI line slopes differently from the original series line, a divergence occurs. Look at the example in Fig.4: two local maxima follow two different trends sloped oppositely. In the case shown, the analyst will interpret this divergence as a bullish expectation (since the RSI oscillator diverges from the original series: it starts increasing when the original series is still decreasing). In our simplified model, the presence of such a divergence translates into a change in the prediction of the $I(t+1)-I(t)$ sign, depending on the bullish or bearish trend of the previous $T_{RSI}$ days. In the following simulations we will choose $\tau_{RSI}=T_{RSI}=14$ days, since - again - this value is one of the mostly used in RSI-based actual trading strategies. See Ref. \cite{Murphy}.  

{\it 4) Up and Down Persistency (UPD) Strategy}
\\
This deterministic strategy does not come from technical analysis. However, we decided to consider it because it seems to follows the apparently simple alternate "up and down" behavior of market series that any observer can see at first sight. The strategy is based on the following very simple rule: the prediction  for tomorrow market's behavior is just the opposite of what happened the  day before. If, e.g., one has $I(t)-I(t-1)>0$, the expectation at time $t$ for the period $t+1$ will be bullish: $I(t+1)-I(t)<0$, and vice versa. 

{\it 5) Moving Average Convergence Divergence (MACD) Strategy}
\\
The 'MACD' is a series built by means of the difference between two Exponential Moving Averages (EMA, henceforth) of the market price, referred to two different time windows, one smaller and one larger. In any moment {\it t}, $MACD_{t}=EMA^{12d}_{t}-EMA^{26d}_{t}$. In particular, the first is the Exponential Moving Average of $I(t)$ taken over twelve days, whereas the second refers to twenty-six days. The calculation of these EMAs on a pre-determined time lag, {\it x}, given a proportionality weight $w=\frac { 2 }{ x+1 } $, is executed by the following recursive formula: $EMA^{xd}_{t}=EMA^{xd}_{t-1}+w[I(t)-EMA^{xd}_{t-1}]$ with $EMA^{ xd }_{ 0 }=\frac { \sum _{ j=t-x }^{ t }{ I(j) }  }{ x }$, where $t\ge x$. Once the MACD series has been calculated, its 9-days Exponential Moving Average is obtained and, finally, the trading strategy for the market dynamics prediction can be defined: the expectation for the market is bullish (bearish) if $MACD-EMA_{MACD}^{9d}>0$ ($MACD-EMA_{MACD}^{9d}<0$). See Ref. \cite{Murphy}.

\section{Results of Empirically Based Simulations}

For each one of our four financial time series of length $T$ (in days), the goal was simply to predict, day by day and for each strategy, the upward (bullish) or downward (bearish) movement of the index $I(t+1)$ at a given day with respect to the closing value $I(t)$ one day before: if the prediction is correct, the trader wins, otherwise he/she looses. In this connection we are only interested in evaluating the percentage of wins achieved by each strategy, assuming that - at every time step - the traders perfectly know the past history of the indexes but do not possess any other information and can neither exert any influence on the market, nor receive any information about future moves.  
\\
In the following, we test the performance of the five strategies by dividing each of the four time series into a sequence of $N_w$ trading windows of equal size $T_w=T/N_w$ (in days) and evaluating the average percentage of wins for each strategy inside each window while the traders move along the series day by day, from $t=0$ to $t=T$. This procedure, when applied for $N_w = 3, 9, 18, 30$, allows us to explore the performance of the various strategies for several time scales (ranging, approximatively, from $6$ months to $5$ years). 
\\
\begin{figure}  
\begin{center}
\includegraphics[width=3.5in,angle=0]{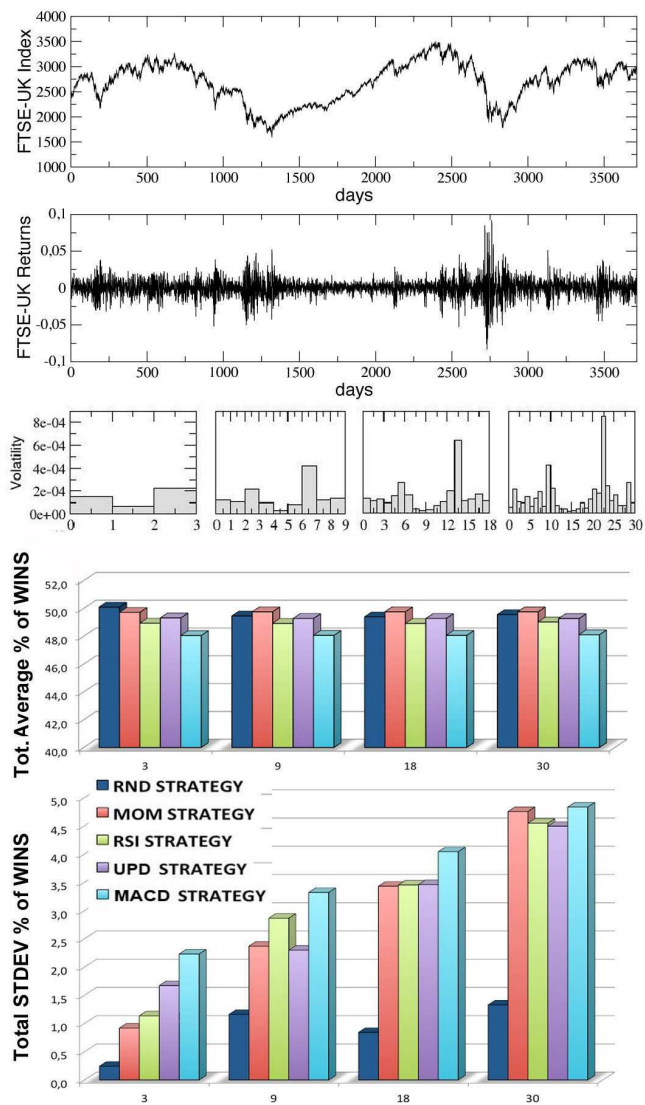}
\end{center}
\caption{{\bf Results for the FTSE-UK index series, divided into an increasing number of trading-windows of equal size ($3,9,18,30$), simulating different time scales.} From top to bottom, we report the index time series, the corresponding returns time series, the volatility, the percentages of wins for the five strategies over all the windows and the corresponding standard deviations. The last two quantities are averaged over $10$ different runs (events) inside each window.}
\end{figure}
\begin{figure}  
\begin{center}
\includegraphics[width=3.5in,angle=0]{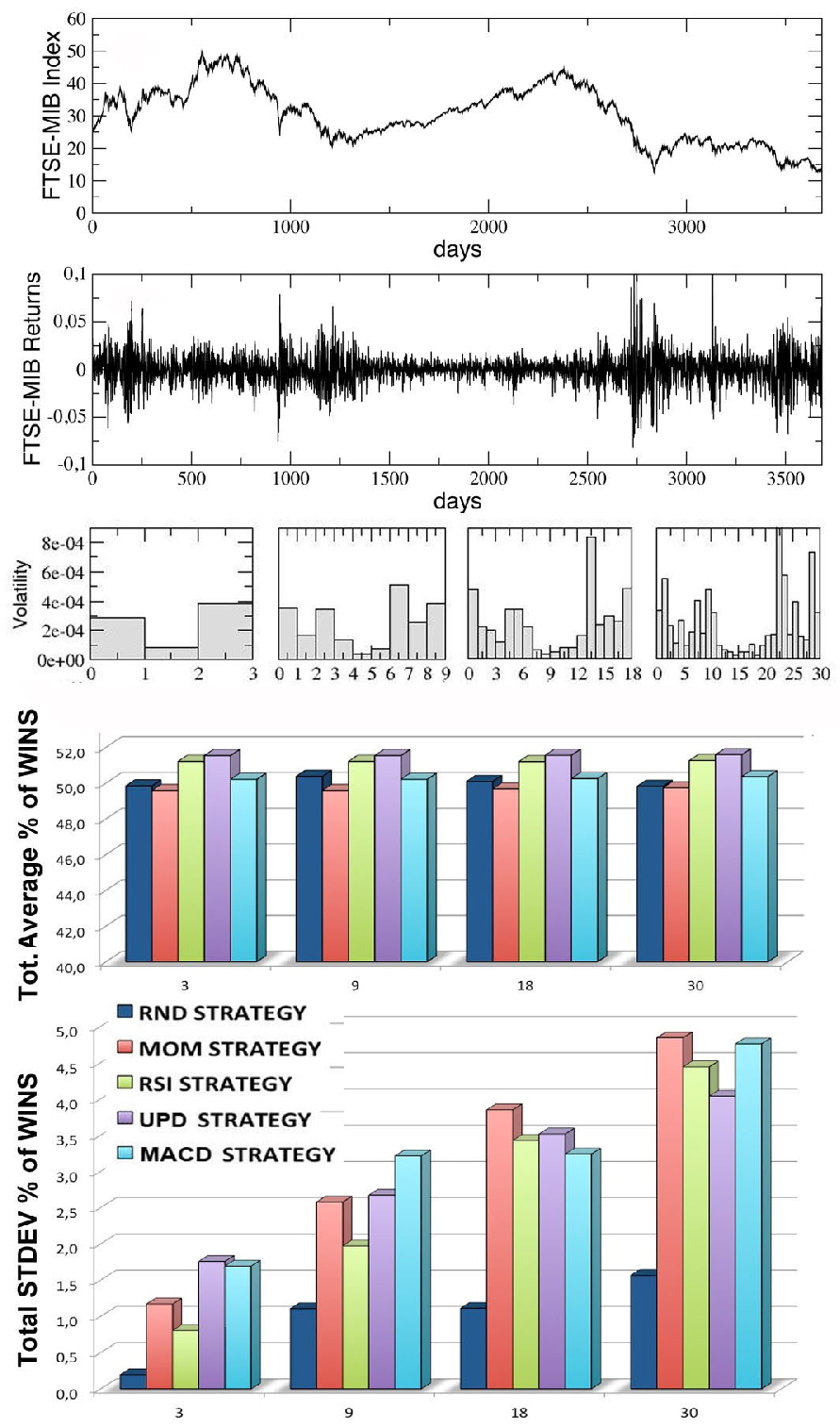}
\end{center}
\caption{{\bf Results for the FTSE-MIB index series, divided into an increasing number of trading-windows of equal size ($3,9,18,30$), simulating different time scales.} From top to bottom, we report the index time series, the corresponding returns time series, the volatility, the percentages of wins for the five strategies over all the windows and the corresponding standard deviations. The last two quantities are averaged over $10$ different runs (events) inside each window.}
\end{figure}
\begin{figure}  
\begin{center}
\includegraphics[width=3.5in,angle=0]{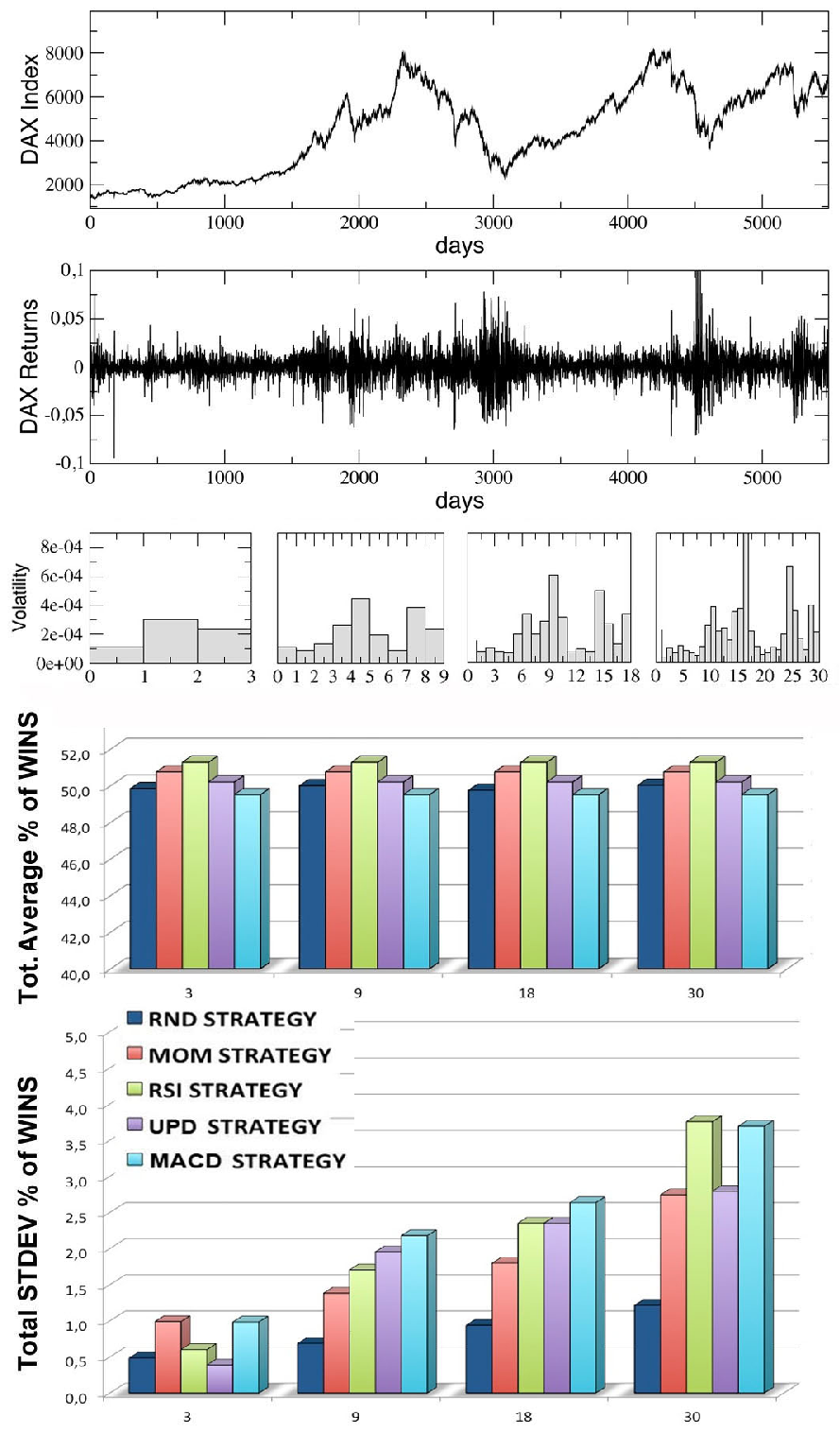}
\end{center}
\caption{{\bf Results for the DAX index series, divided into an increasing number of trading-windows of equal size ($3,9,18,30$), simulating different time scales.} From top to bottom, we report the index time series, the corresponding returns time series, the volatility, the percentages of wins for the five strategies over all the windows and the corresponding standard deviations. The last two quantities are averaged over $10$ different runs (events) inside each window.}
\end{figure}
\begin{figure}  
\begin{center}
\includegraphics[width=3.5in,angle=0]{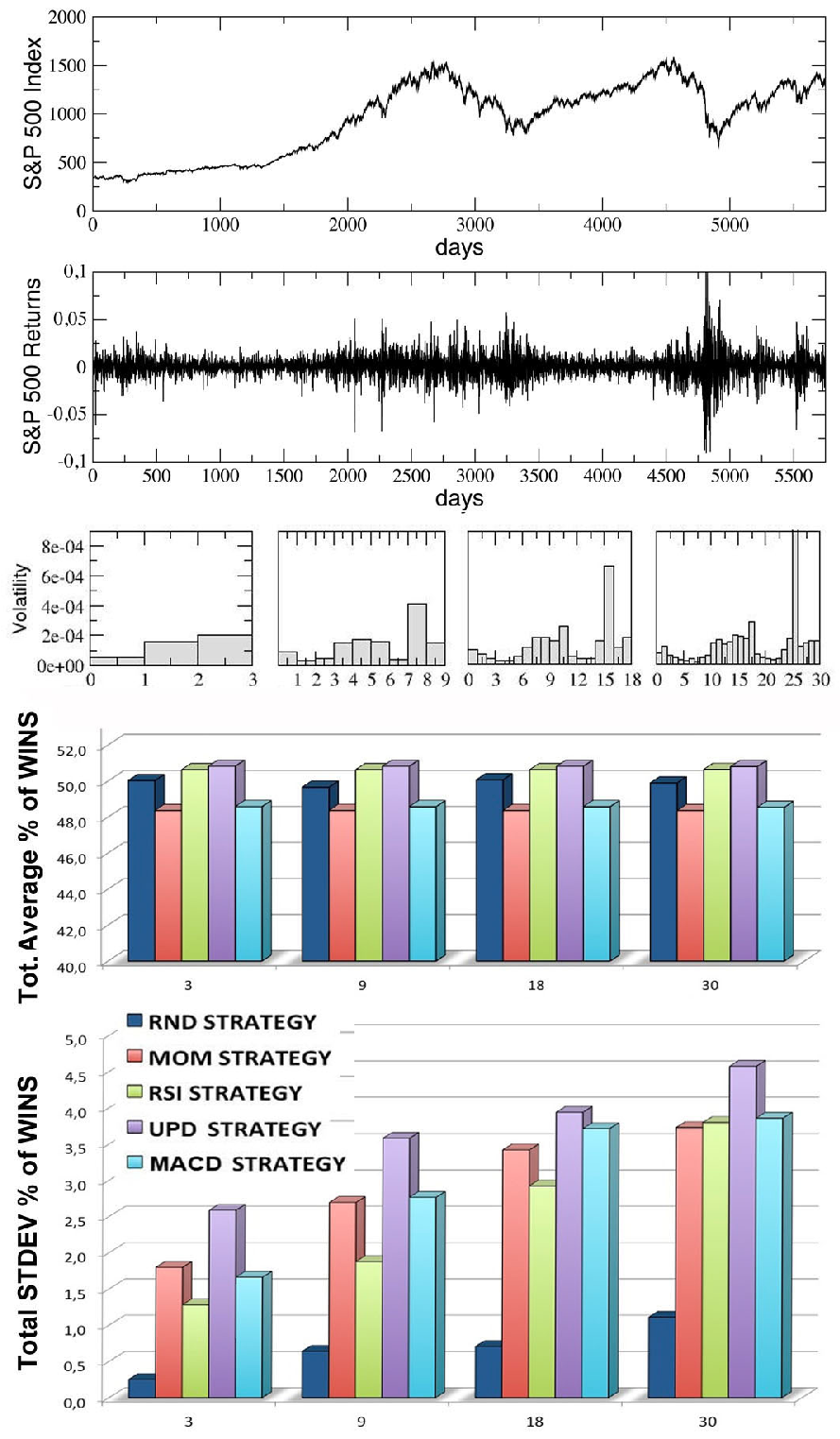}
\end{center}
\caption{{\bf Results for the S$\&$P 500 index series, divided into an increasing number of trading-windows of equal size ($3,9,18,30$), simulating different time scales.} From top to bottom, we report the index time series, the corresponding returns time series, the volatility, the percentages of wins for the five strategies over all the windows and the corresponding standard deviations. The last two quantities are averaged over $10$ different runs (events) inside each window.}
\end{figure}
The motivation behind this choice is connected to the fact that the time evolution of each index clearly alternates between calm and volatile periods, which at a finer resolution would reveal a further, self-similar, alternation of intermittent and regular behavior over smaller time scales, a characteristic feature of turbulent financial markets \cite{mandelbrot,mandelbrot2,mantegna-stanley,holyst}. 
Such a feature makes any long-term prediction of their behavior very difficult or even impossible with instruments of standard financial analysis. The point is that, due to the presence of correlations over small temporal scales (as confirmed by the analysis of the time dependent Hurst exponent in Fig. 3), one might expect that a given standard trading strategy, based on the past history of the indexes, could perform better than the others inside a given time window. But this could depend much more on chance than on the real effectiveness of the adopted algorithm. On the other hand, if on a very large temporal scale the financial market time evolution is an uncorrelated Brownian process (as indicated by the average Hurst exponent, which result to be around $0.5$ for all the financial time series considered), one might also expect that the performance of the standard trading strategies on a large time scale becomes comparable to random ones. In fact, this is exactly what we found as explained in the following.
\\
In Figs. 5-8, we report the results of our simulations for the four stock indexes considered (FTSE-UK, FTSE-MIB, DAX, S$\&$P 500). In each figure, from  top to  bottom, we plot: the market time series $I(t)$ as a function of time; the correspondent 'returns' series, determined as the ratio $[I(t+1)-I(t)]/I(t)$; the volatility of the returns, i.e. the variance of the previous series, calculated inside each window for $4$ increasing values of  the trading window size $N_w$ (equal to, from left to right, $3$, $9$, $18$ and $30$ respectively); the average percentage of wins for the five trading strategies considered, calculated for the same four kinds of windows (the average is performed over all the windows in each configuration, considering $10$ different simulation runs inside each window); the corresponding standard deviations for the wins of the five strategies.     
\\
Observing the last two panels in each figure, two main results are evident:
\\
\begin{figure}  
\begin{center}
\includegraphics[width=3.5in,angle=0]{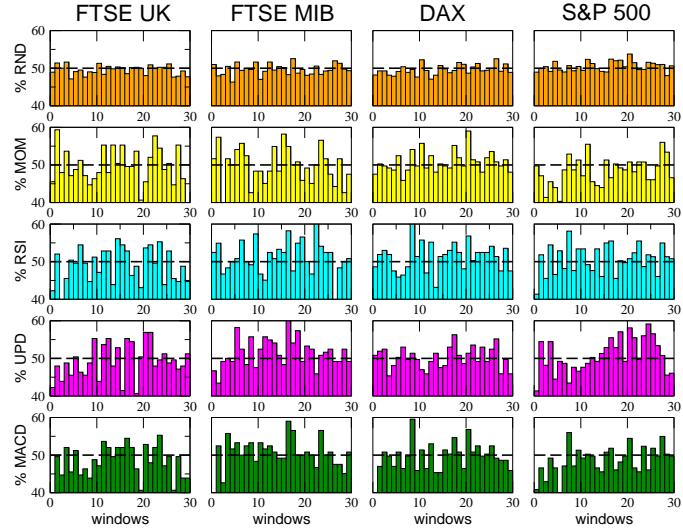}
\end{center}
\caption{{\bf The percentage of wins of the different strategies inside each time window - averaged over $10$ different events - is reported, in the case $N_w=30$, for the four markets considered.} As visible, the performances of the strategies can be very different one from the others inside a single time window, but averaging over the whole series these differences tend to disappear and one recovers the common $50\%$ outcome shown in the previous figures.
}
\end{figure}
%
1. The average percentages of wins for the five strategies are always comparable and oscillate around $50\%$, with small random differences which depend on the financial index considered. The performance of $50\%$ of wins for all the strategies may seem paradoxical, but it depends on the averaging procedure over all the windows along each time series. In Fig. 9 we show, for comparison, the behavior of the various strategies for the four financial indexes considered and for the case $N_w=30$ (the score in each window is averaged over $10$ different events): as one can see, within a given trading window each single strategy may randomly perform much better or worse than $50\%$, but on average the global performance of the different strategies is very similar. Moreover, referring again to Figs. 5-8, it is worth to notice that the strategy with the highest average percentage of wins (for most of the windows configurations)  changes from one index to another one: for FTSE-UK, the MOM strategy seems to have a little advantage; for FTSE-MIB, the UPD seems to be the best one; for DAX, the RSI, and for the S$\&$P 500, the UPD performs slightly better than the others. In any  case the advantage of a strategy seems  purely coincidental.     
\\
2. The second important result is that the fluctuations of the random strategy are always smaller than those of the other strategies (as it is also visible in Fig. 9 for the case $N_w=30$): this means that the random strategy is less risky than the considered standard trading strategies, while the average performance is almost identical. This implies that, when attempting to optimize the performance, standard traders are fooled by the "illusion of control" phenomenon \cite{Satinover1,Satinover2}, reinforced by a lucky sequence of wins in a given time window. However, the first big loss may drive them out of the market. On the other hand, the effectiveness of random strategies can be probably related to the turbulent and erratic character of the financial markets: it is true that a random trader is likely to win less in a given time window, but he/she is likely also to loose less. Therefore his/her strategy implies less risk, as he/she has a lower probability to be thrown out of the game.

\section{Conclusions and Policy Implications}
     
In this paper we have explored the role of random strategies in financial systems from a micro-economic point of view. In particular, we simulated the performance of five trading strategies, including a completely random one, applied to four very popular financial markets indexes, in order to compare their predictive capacity. Our main result, which is  independent of the market considered, is that standard trading strategies and their algorithms, based on the past history of the time series, although have occasionally the chance to be successful inside small temporal windows, on a large temporal scale  perform on average not better than the purely random strategy, which, on the other hand, is also much less volatile. In this respect, for the individual trader, a purely random strategy represents a costless alternative to expensive professional financial consulting, being at the same time  also  much less risky, if  
compared to the other trading strategies.
\\ 
This result, obtained at a micro-level, could have many implications for real markets also at the macro-level, where other important phenomena, like herding, asymmetric information, rational bubbles occur.
In fact, one might expect that a widespread adoption of  a random approach for financial transactions would result in a more stable market with lower volatility. In this connection, random strategies could play the role of reducing herding behavior over the whole market since, if agents knew that financial transactions do not necessarily carry an information role, bandwagon effects could  probably fade. 
On the other hand, as recently suggested by one of us \cite{helbingchristen}, if the policy-maker (Central Banks) intervened by randomly buying and selling financial assets, two results could be simultaneously obtained. From an individual point of view, agents would suffer less for asymmetric or insider information, due to the consciousness of a "fog of uncertainty" created by the random investments. From a systemic point of view, again the herding behavior would be consequently reduced and eventual bubbles would burst when they are still small and are less dangerous; thus, the entire financial system would be less prone to the speculative behavior of credible "guru" traders, as explained also in \cite{gallegatietal}. Of course, this has to be explored in detail as well as the feedback effect of a global reaction of the market to the  application of these actions.This topic is however beyond the goal of the present paper and it will be investigated in a future work.   

\section*{Acknowledgments}
We thank H. Trummer for DAX historical series and the others institutions for the respective data sets.


\end{document}